\documentclass[preprint,showpacs,preprintnumbers,amsmath,amssymb,prl]{revtex4}

\usepackage{graphicx}
\usepackage{dcolumn}
\usepackage{bm}

\begin{document}

\title{Formation and transportation of sand-heap in an inclined and vertically
 vibrated container}

\author{Guoqing Miao}
\email{miaogq@nju.edu.cn}
\author{Kai Huang}
\author{Yi Yun}
\author{Peng Zhang}
\author{Weizhong Chen}
\author{Xinlong Wang}
\author{Rongjue Wei}

\affiliation{ State Key Laboratory of Modern Acoustics and
Institute of Acoustics, Nanjing University, Nanjing 210093, P. R.
China }

\date{January 1, 2005}

\begin{abstract}
We report the experimental findings of formation and motion of
heap in granular materials in an inclined and vertically vibrated
container. We show experimentally how the transport velocity of
heap up container is related to the driving acceleration as well
as the driving frequency of exciter. An analogous experiment was
performed with a heap-shaped Plexiglas block. We propose that
cohesion force resulted from pressure gradient in ambient gas
plays a crucial role in enhancing and maintaining a heap, and
ratchet effect causes the movement of the heap. An equation which
governs the transport velocity of the heap is presented.
\end{abstract}

\maketitle

It is well known that under vibration many processes, e.~g.
segregation, convection, heaping, density wave, and anomalous
sound propagation, which govern the physics of granular materials,
are quite unusual~\cite{Jaeger}, so that the properties of such
materials are not well understood. For example, heaping is one of
long-standing problem since Faraday~\cite{Faraday, Duran, Evesque,
Thomas}, and several physical mechanisms have been identified as
possible causes of it: friction between the walls and
particles~\cite{Clement}, analog of acoustic streaming if the
shaking is nonuniform~\cite{Savage}, gas pressure
effect~\cite{Pak}, and auto-amplification~\cite{Chen}. Recently we
have observed experimentally the formation of a heap and the
motion of it from lower to higher end of an inclined and
vertically vibrated container. The experiment was conducted in a
Plexiglas rectangular container [370 $mm$ (length) $\times$25 $mm$
(width) $\times$80 $mm$ (height)]. We investigated the behavior of
two types of quartz sands: spheres of diameter 0.15-0.20 $mm$ and
grains of irregular shape or coarse surface with diameter 0.3-0.5
$mm$. The container was inclined with an inclination $\alpha$ from
0.04 radian to 0.25 radian by putting a pad underneath it. The
vibration exciter (Br\"{u}el \& Kj$\ae$r 4805) was driven by a
sinusoidal signal, and controlled by a vibration exciter control
(Br\"{u}el \& Kj$\ae$r 1050). Driving frequency $f$ and
dimensionless acceleration amplitude $\Gamma=4\pi^2f^2A/g$ (where
$A$ is driving amplitude and $g$ the gravitational acceleration)
were used as two control parameters.

\begin{figure}[b]
\includegraphics [width=0.45\textwidth]{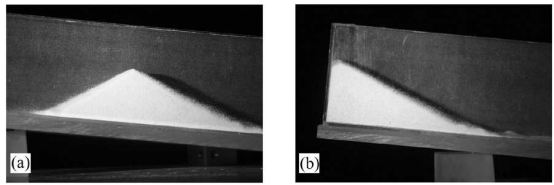}
\caption{\label{fig:1} The center-high heap (a) and the wall-high
heap (b) of coarse sands. $\alpha=0.087$ radian, $f=15$ Hz,
$\Gamma=2$.}
\end{figure}
The experiment shows that even if in horizontal container the
horizontal acceleration can cause movement of a heap. The
horizontal acceleration of our exciter is about 2.5\% of the
vertical acceleration. To get rid of the influence of the
horizontal component of the driving acceleration on the movement
of the heap along the longer direction of the container, we
adjusted the longer direction of the container orthogonal to
horizontal acceleration of the exciter. The ranges of $\Gamma$ and
$f$ we used were from 1.4 to 2.8 and from 11 Hz to 20 Hz
respectively, which are good ranges for heap formation. At first,
about 80~$ml$ of sands was uniformly put into the lower part of
the vessel. Then as $\Gamma$ increased to and beyond some critical
acceleration $\Gamma_c (>1)$, center-high heaps formed, meanwhile
they moved up the container. As center-high heaps reached higher
end wall of the container, they moved forward continuously until
they became wall-high heaps.
\begin{figure*}[t]
\includegraphics[width=0.8\textwidth]{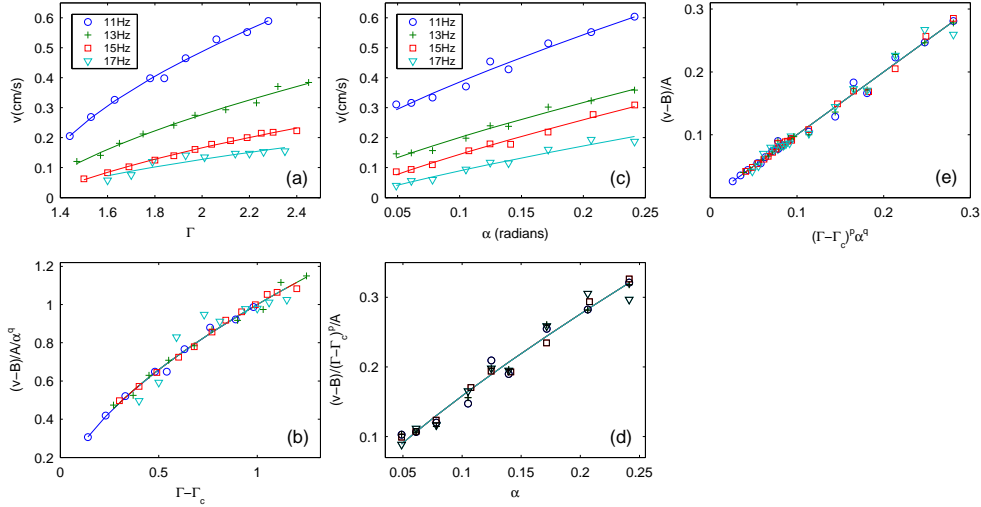}
\caption{\label{fig:2} The transport velocity $V$ of heap formed
by coarse sands. (a) $V$ vs $\Gamma$ for different frequencies.
The solid lines are fits by $V=A(\Gamma-\Gamma_c)^p\alpha^q+B$
with $\alpha=0.045$ radian. (b) $(V-B)/(A\alpha^q)$ vs
$\Gamma-\Gamma_c$ for the data in (a). The solid line is
$(\Gamma-\Gamma_c)^p$. (c) $V$ vs $\alpha$ for different
frequencies. The solid lines are fits by
$V=A(\Gamma-\Gamma_c)^p\alpha^q+B$ with $\Gamma=2$. (d)
$(V-B)/[A(\Gamma-\Gamma_c)^p]$ vs $\alpha$ for the data in (c).
The solid line is $\alpha^q$. (e) $(V-B)/A$ vs
$(\Gamma-\Gamma_c)^p\alpha^q$ for the data both in (a) and (c).
The solid line is $(\Gamma-\Gamma_c)^p\alpha^q$.}
\end{figure*}
Fig.~1 shows the photos of heaps formed by coarse sands. Fig.~1(a)
is center-high heap, and (b) is wall-high heap. For center-high
heap, the back (or right shown as in the figure) surface is longer
than the frontal (or left) one, but the dynamical angles of repose
(slightly smaller than the maximum angle of repose of the static
heap) of both frontal and back surfaces relative to horizontal are
the same. The difference between the heaps formed by two types of
sands is: the dynamical angle of repose of the heap formed by
coarse sands is larger than that formed by spherical sands. The
heaps moved up the container with nearly uniform velocity. We
measured the velocities $V$ of the heaps. Fig.~2 gives the results
for coarse sands. Fig.~2(a) shows $V$ vs $\Gamma$ for different
$f$. One can see that $V$ increases with $\Gamma$ for all of
frequencies, but decreases as $f$ increases for all values of
$\Gamma$. The velocity of heap formed by coarse sands is greater
than that of heap formed by spherical sands for all sets of values
of $\Gamma$ and $f$. Fig.~2(c) shows the velocity of heap as a
function of inclination of the container. It is shown that $V$
increases as $\alpha$, and behaves in the same way as in Fig.~2(a)
as $f$ changes. The heap formed by coarse sands moved faster than
the heap formed by spherical sands did for the same inclination
and the same driving parameters.

Obviously, the transport of the heap is a cooperative behavior of
the granular materials. To examine this idea we put a Plexiglas
block the same in shape and dimension as the sand heap on the same
vibrated inclined container, a similar transport of the block up
the container was observed. So we consider that the transport of
the heap is similar to that of a solid block. But why does it
move, or what is the mechanism of the transport? We used a high
speed camera (Redlake MASD MotionScope PCI 2000~SC) to record the
movement of the block as it moved up the container with record
rate of 250 fps (frames per second), then played back slowly (25
fps). In this way, we can see the detail of the movement of the
block. We marked the center of mass of the block with a black
point C (shown as in Fig.~4). Fig.~3(a) is the orbit of the center
of mass of the block measured experimentally in laboratory
reference frame. Fig.~3(b) is a schematic diagram of Fig.~3(a),
with which we can discuss the movement of the block conveniently.
To examine the effect of ambient gas, we performed two things.
\begin{figure}[b]
\includegraphics [width=0.35\textwidth] {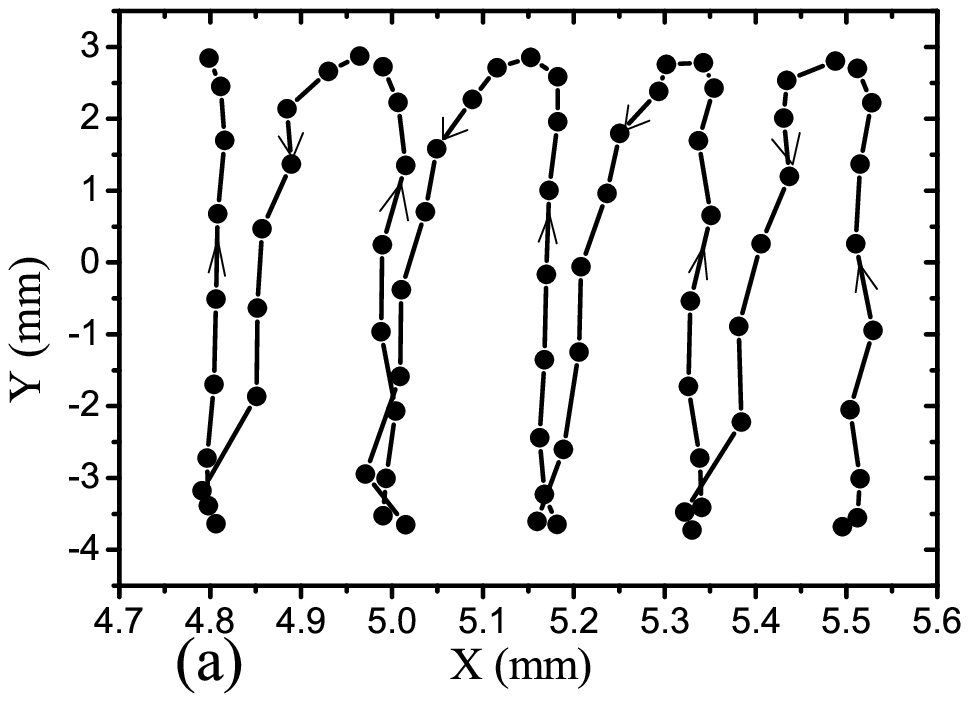}
\includegraphics [width=0.3\textwidth] {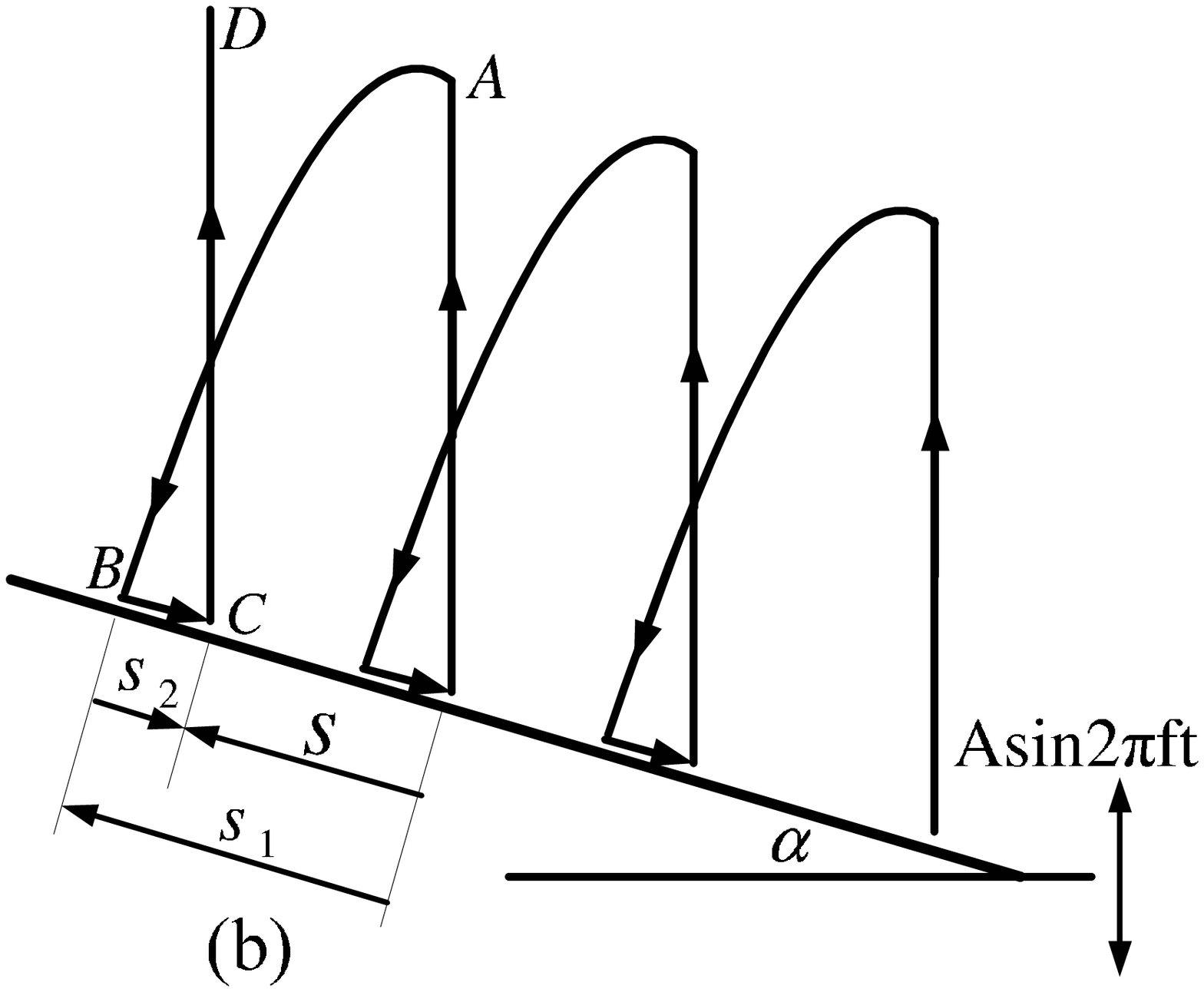}
\caption{\label{fig:3} (a) The orbit of the center of mass of the
block measured experimentally. $\alpha=0.06$ radian, $f=15$ Hz,
$\Gamma=2.3$. (b) The schematic diagram of (a).}
\end{figure}
First, to get rid of the pressure difference between the gap and
the above of the block, we drilled a number of holes vertically
and parallel each other through the block, and observed no
transport of the block. And second, we evacuated the air from the
container and also observed no transport of the block (either with
or without holes drilled through the block). These imply that air
pressure difference plays a critical role for the movement of the
block up the container. Therefore we propose a mechanism for the
movement of the block up the container as follows. In each cycle,
when $\Gamma\cos 2\pi ft<-g$, block separates with container, and
a gap forms between block bottom and the container floor. The
pressure $p_1$ in gap is less than the atmospheric pressure $ p_0$
above the block on an average (ref.~\cite{Chlenov} shows that the
mean pressure in a gap between the bottom of granular bed and the
container floor is below atmospheric pressure. And as Faraday
Said: in this gap, ``it forms a partial vacuum''. We consider this
also suits to the case of block). This pressure difference causes
a force exerting on the block. Fig.~4 is a schematic diagram for
the forces acting on the block (or heap) during free flight. The
forces acting on the frontal (I) and back (II) parts (with dashed
line as intersection) are represented by vectors $\bf F_{1}$ and
$\bf F_{2}$ at the centers of mass of two parts, ${\rm c}_1$ and
${\rm c}_2$, respectively. They are perpendicular to frontal and
back surface (i.~e. the frontal and back sides in the figure),
respectively.
\begin{figure}[b]
\includegraphics [width=0.4\textwidth] {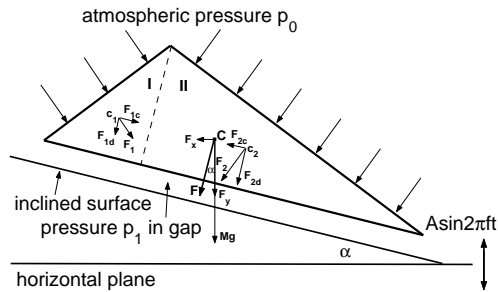}
\caption{\label{fig:4} A schematic diagram for the forces acting
on block (heap) during free flight.}
\end{figure}
The magnitudes of them are proportional to lengths of the frontal
and back sides, respectively. They are decomposed into two
components: those parallel (represented by $\bf F_{1c}$ and $\bf
F_{2c}$) and those perpendicular (represented by $\bf F_{1d}$ and
$\bf F_{2d}$) to the bottom of the block, respectively. The
parallel components $\bf F_{1c}$ and $\bf F_{2c}$ equal in
magnitude but opposite in direction. So the total force $\bf F$ is
only the sum of perpendicular components $\bf F_{1d}$ and $\bf
F_{2d}$, acts at the center of mass of the block, $C$, is
perpendicular to the bottom of block, and makes an angle $\alpha$
(i.~e. the inclination of the container) with gravitational force
of the block. This force, together with gravitation force $\bf Mg$
($\bf M$ is mass of the block), force the block to move along a
ballistic trajectory until collides with the container
[$A\rightarrow B$ in Fig.~3(b)]. Upon colliding with the
container, the block slides down the container [$B\rightarrow C$
in Fig.~3(b)] due to gravity, meanwhile moves together with the
container in vertical direction until next separation with
container [$C\rightarrow D$ in Fig.~3(b)]. Then a new cycle
begins. But due to the friction between the block and the
container floor, the distance up the container during free flight
is much larger than the distance down the container during
bed-floor collision, i.~e. the block moves one step up the
container in each cycle. The analysis above shows that the
transport of block up the container is a ratchet effect caused by
the pressure difference (between the gap and above of the block)
and the friction force (between block and the container floor).
\begin{figure}[b]
\includegraphics [width=0.45\textwidth]{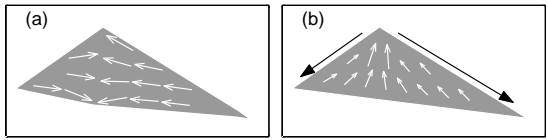}
\caption{\label{fig:5} The schematic diagram of the velocity (or
mass flow) field in a heap formed by coarse sands during the
period of free flight (a) and heap-floor collision(b),
respectively.}
\end{figure}

For the granular bed, if for any reason, some small (or flatter)
initial heap has formed. In the period of free flight, the
pressure difference between the gap and above of the
heap~\cite{Chlenov} leads to a pressure gradient (or force) in the
interior of the heap. Also cf. Fig.~4, and we still use $\bf F_1$
and $\bf F_2$ as representatives of the total force acting on
frontal and back parts of the heap, respectively. These two forces
are perpendicular to the frontal and back surface, respectively.
Here we call the parallel components $\bf F_{1c}$ and $\bf F_{2c}$
as a couple of cohesion force, i.~e., they make granular materials
cohesive and enhance the heap. Under the action of the cohesion
force, heap is compressed in direction parallel to heap bottom
while elongated in direction perpendicular to heap bottom. And
this makes the heap bottom convex, as we have observed
experimentally in both inclined and horizontal
containers~\cite{Chen}. Fig.~5(a) is a schematic diagram of the
velocity (or mass flow) field in the interior of the heap formed
by coarse sands during free flight. The total force $\bf F$ acting
on the center of mass of heap, together with gravitational force
$\bf Mg$, force the heap as a whole to move along a ballistic
trajectory and up the container. Upon colliding with the
container, the center part of the heap bottom touches floor first,
then the other parts, from center to outer, touch floor
consecutively. This results in a further enhancement of the heap.
In this way the slope of the heap is getting larger and larger.
Once the slope angle reaches and exceeds the dynamic angle of
repose of the heap, and upon colliding with container, the
avalanche occurs on the surface of the heap. Then the process
repeats periodically: during the free flight, the compression due
to cohesion force makes the slope of heap exceeding the dynamic
angle of repose, and during the bed-floor collision, avalanche
occurs. So when the heap reaches a steady state, in the laboratory
reference frame, one can see a steady convection flow: in the
interior of the heap, grains move upward, while at the surface
grains move rapidly downward [schematically shown as in
Fig.~5(b)]. This analysis is also suitable to wall-high heap, in
which the cohesion force points higher end wall. Similar to the
block, the transport of the heap up the container is also a
ratchet effect, which is caused by the pressure gradient in the
heap and the friction force between the heap and the container
floor. If we pump the air out of the container, as pressure is
reduced, heap reaches a flat state gradually. This shows that
ambient gas plays a important role in enhancing and maintaining a
heap.

Let us now describe the ratchet effect little more concretely but
qualitatively [also cf. Fig.~3(b) and Fig.~4]. We denote $\beta$
as the ratio of free-flight time to the excitation period $T$ in
each cycle. In each cycle, the up-container distance of free
flight by heap is
$s_1\sim a_1\beta^2T^2/\cos\alpha$,
where $\alpha$ is the inclination of the container, and $a_1$ is
the mean acceleration of the center of mass of the heap due to the
horizontal component $\bf F_x$ of the force $\bf F$. Then $a_1\sim
F\sin\alpha$ $(F=|\bf F|)$, and $s_1\sim F\beta^2T^2\tan\alpha$.
In period of bed-floor collision, heap slides down the container.
The distance of sliding is
$s_2\sim a_2(1-\beta)^2T^2$,
where $a_2$ is the mean acceleration down the container, which is
determined by the friction coefficient $\mu$ between the granular
bed and the container floor, the inclination $\alpha$ of the
container, and gravitational acceleration $g$ through
$a_2=g\sin\alpha-\mu g\cos\alpha$.
The total (or net) displacement of heap up the container in each
cycle is $s=s_1-s_2$, timing driving frequency $f$ gives the
velocity $V$ of heap up the container. The ratio $\beta$ increases
with driving amplitude $A=\Gamma g/4\pi^2f^2$~\cite{Miao}, i.~e.
$V$ increases as $\Gamma$, but decreases as $f$ increases. The
larger the $\mu$, the smaller the $s_2$, and then the lager the
$V$. Because the friction between coarse sands and the container
floor is larger than that between spherical sands and the
container floor, the transport velocity of heap formed by coarse
sands is larger than that of heap formed by spherical sands. In
view of $s_2<<s_1$ and $\alpha<<1$ in our experiment, then $V$
increases as inclination $\alpha$ of the container. The pressure
gradient or force in the interior of heap depends on driving
acceleration and frequency, inclination of the container, and
properties of grains in a specific and complex way, and changes
with time and space. The investigation on this problem is under
way.

Any way, from the description above we have already had a
qualitative picture on how the transport velocity of the heap up
the container depends on driving acceleration and frequency,
inclination of the container, properties of the grains, and
friction between heap and container floor. Based on this and on
the analysis of experimental results, the experimental data in
Fig.~2(a) and (c) are well fitted by the equation
\begin{equation}
V=A(\Gamma-\Gamma_c)^p\alpha^q+B, 
\end{equation}
with $p=0.6$, $q=0.8$, and $\alpha=0.045$ radian for Fig~2(a) and
$\Gamma=2$ for Fig~2(c), respectively. The solid lines in the
figure are nonlinear least-square fits by Eq.~(1). The parameters
$A$, $B$ and $\Gamma_c$ are depend on the driving frequency $f$,
the dynamic angle of repose of heap which depends on properties
(e.~g. size and shape etc.) of grains, and the friction
coefficient $\mu$ between granular bed and container floor. If the
data of Fig.~2(a) are plotted as $(V-B)/(A\alpha^q)$ vs
$\Gamma-\Gamma_c$, they collapse onto a single curve, Fig.~2(b);
and if the data of Fig.~2(c) are plotted as
$(V-B)/[A(\Gamma-\Gamma_c)^p]$ vs $\alpha$, they also collapse
onto a single curve, Fig.~2(d), within the experimental
resolution. Moreover, if the data of both Fig.~2(a) and Fig.~2(c)
are plotted as $(V-B)/A$ vs $(\Gamma-\Gamma_c)^p\alpha^q$, they
all collapse onto a single curve, Fig.~2(e).

This model is also suitable for the heap (either the center-high
or the wall-high heap) formed in a horizontal container, where the
cohesion force (as above) plays the same role (enhancing and
maintaining the heap) as in an inclined container. The total force
$\bf F$ due to the pressure gradient is perpendicular to heap
bottom, i.~e. is parallel to gravitational force $\bf Mg$, and no
force forces heap to move in horizontal direction.

Our conclusion is: pressure gradient in ambient gas, which makes
cohesionless granular materials cohesive, plays a crucial role in
enhancing and maintaining a heap in vibrating granular materials;
pressure gradient in ambient gas and friction force between the
granular bed and the container floor, which lead to a ratchet
effect, are unique cause for the transport of a heap up an
inclined container. The transport velocity of the heap is well
described by the equation (1). Our mechanism for enhancing and
maintaining heap also shows that one reason for lacking of heaping
in molecular dynamics (MD) simulation may be that those models did
not take into account the cohesion force due to ambient gas
effect.

This work was supported by the Special Funds for Major State Basic
Research Projects, National Natural Science Foundation of China
through Grant Nos. 10074032 and 10474045, and the Research Fund
for the Doctoral Program of Higher Education of China through
Grant No. 20040284034.\\

\end{document}